\documentclass{jaa}

\usepackage{graphicx}
\usepackage{gensymb}
\usepackage{authblk}
\usepackage[normalem]{ulem}

\begin{document}\sloppy

\title{Observations of AR Sco with $Chandra$ and  {\it AstroSat} Soft X-ray Telescope}


\author{K. P. Singh\textsuperscript{1,*}, V. Girish \textsuperscript{2}, J. Tiwari\textsuperscript{1}, P. E. Barrett \textsuperscript{3}, D. A. H. Buckley\textsuperscript{4}, S. B. Potter \textsuperscript{4,8},  E. Schlegel \textsuperscript{5}, V. Rana  \textsuperscript{6} and G. Stewart  \textsuperscript{7}}
\affilOne{\textsuperscript{1}Indian Institute of Science Education and Research Mohali, SAS Nagar, Sector 81, P.O.Manauli- 140306, India.\\}
\affilTwo{\textsuperscript{2}Indian Space Research Organisation HQ, New BEL Road, Bengaluru 560094, India.\\}
\affilFour{\textsuperscript{3} The George Washington University, 725 21st St. NW, Washington, DC 20052\\}
\affilThree{\textsuperscript{4}South African Astronomical Observatory, PO Box 9, Observatory Road, Observatory 7935, Cape Town, South Africa.\\}
\affilFive{\textsuperscript{5} The University of Texas at San Antonio,  One UTSA Circle, San Antonio, TX  78249, USA\\}
\affilSix{\textsuperscript{6} Raman Research Institute, C. V. Raman Avenue, Sadashivanagar,  Bengaluru 560 080, India\\}
\affilSeven{\textsuperscript{7} Department of Physics and Astronomy, The University of Leicester, University Road, Leicester, LE1 7RH, UK\\}
\affilEight{\textsuperscript{8} Department of Physics, University of Johannesburg, PO Box 524, Auckland Park 2006, South Africa}


\twocolumn[{

\maketitle

\corres{kpsinghx52@gmail.com}

\msinfo{xxxx}{xxxx}

\begin{abstract}
We present  our {\it AstroSat} soft X-ray observations of a compact binary system, AR Sco, and analysis of its X-ray observations with $Chandra$ that were taken only about a week before the {\it AstroSat} observations. An analysis of the soft X-ray ($0.3-2.0$ keV) data limits the modulation of the spin, orbital, or beat periods to less than 0.03 counts s$^{-1}$ or $<$10\% of the average count rate. The X-ray flux obtained from both observatories is found to be almost identical (within a few percent) in flux, and about 30\% lower than reported from the nine months older observations with {\it XMM-Newton}.  A two-temperature thermal plasma model with the same spectral parameters fit $Chandra$ and {\it AstroSat} data very well, and requires very little absorption in the line of sight to the source.  The low-temperature component has the same temperature ($\sim$1 keV) as reported earlier, but the high-temperature component has a lower temperature of 5.0$^{+0.8}_{-0.7}$ keV as compared to 8.0 keV measured earlier, however, the difference is not statistically significant.

\end{abstract}
\keywords{Stars: individual: AR  Sco--- X-rays: binaries: close – novae, cataclysmic variables---white dwarfs}

}]


\doinum{12.3456/s78910-011-012-3}
\artcitid{\#\#\#\#}
\volnum{000}
\year{0000}
\pgrange{1--}
\setcounter{page}{1}
\lp{1}
\section{Introduction}
 A uniquely variable star, AR Scorpii (AR Sco) was initially mis-classified as a pulsating $\delta$-Scuti type (Satyvaldiev, V. 1971).   The star is at a distance of 117$\pm$1 pc based on parallax measurements given in {\it Gaia} Early Data Release 3 by Gaia Collaboration, Brown \textit{et al.} (2020), with statistical error as given at the {\it Gaia} website. {\footnote{https://gea.esac.esa.int/archive/.}} It gained prominence when it was identified as a rare radio pulsing white dwarf (WD) binary system by Marsh \textit{et al.} (2016). A very short spin period of P$_s$=117.1 s, and a synodic (spin$-$orbital beat) period, P$_b$=118.2 s and its harmonics, with very high amplitudes of P$_b$ in the ultraviolet, optical and infra-red were found by Marsh \textit{et al.} (2016). These periods were also seen at radio frequencies, and a slowing of the synodic period ($\dot{\rm P}_b$=3.92$\times$10$^{-13}$s s$^{-1}$) was also reported by Marsh et al (2016), thus implying a time-scale of 10$^7$ years for synchronization. However, in more recent photometric studies by Stiller \textit{et al.} (2018) and Gaibor \textit{et al.} (2020), the latter from data taken over a 5 year baseline, a larger spin down rate of $\dot{\rm P}_b$=6.82$\times$10$^{-13}$s s$^{-1}$ is derived, implying a commensurate decrease in the synchonization time. 
  
  Marsh \textit{et al.} (2016) concluded that  AR Sco is a close binary system consisting of a 
  WD and a cool low-mass M-star, with an orbital period of 3.56 h.  They further pointed out that AR Sco is primarily spin-down powered, and its broadband  spectral energy distribution is well explained by synchrotron radiation from relativistic electrons which likely originate from near the WD or are generated in situ at the M star surface through a direct interaction with the WD’s magnetosphere. Marsh \textit{et al.} also reported  on a short X-ray observation of AR Sco using the {\it Swift} X-ray Telescope (XRT)  that showed the X-ray luminosity of AR Sco is very low ($\sim5\times10^{30}$ erg$s^{-1})$ and therefore, accretion is not sufficient to power the system. In some respects AR Scorpii resembles the cataclysmic variable AE Aquarii (Ikhsanov 1998, Meintjes \& Venter 2005, Oruru \& Meintjes 2012), which has a 33 s  white dwarf spin period and a $\dot{\rm P}_b$=5.66$\times$10$^{-14}$s s$^{-1}$ and is in a propeller mass ejection phase, i.e., a system with very little or no mass accretion. However, as pointed out by Marsh \textit{et al.} (2016) and Buckley \textit{et al.} (2017), no flickering or broad emission lines in the optical are seen in  AR Sco, in contrast to AE Aqr. This implies that the system is detached, with little or no mass loss from the secondary star, and therefore by definition is not a cataclysmic variable.  
  
  AR Sco also exhibits strong (up to $\sim40\%$) linear polarization in the optical, modulated strongly  at P$_s$ and P$_b$ (Buckley \textit{et al.} 2017; Potter \& Buckley 2018a), which Buckley \textit{et al.} used to derive an estimate of the  white dwarf's magnetic field of B$\sim$10$^ 8$ Gauss  at its surface. Such a large field was also shown to be consistent with the system being powered by dipole radiation of the slowing down white dwarf, caused by the magnetic torque between the white dwarf and secondary star. A recent model proposed by Lyutikov \textit{et al.} (2020) attempts to explain the properties of AR Sco with a lower white dwarf magnetic field, closer to what is seen in the intermediate polar sub-class of magnetic CVs, which would more easily explain the initial spin-up of the white dwarf. However, the current large spin-down rate, coupled to the very different polarimetric properties of AR Sco compared to intermediate polars, may be argue against a lower field strength. Optical photo-polarimetry and photometry by Potter \& Buckley (2018a,b) has identified several harmonic components of the spin and beat periods, and they have suggested a model in which the optical polarized emission comes from two diametrically opposed synchrotron emission regions in the magnetosphere of the WD, consistent with the model proposed by Takata \textit{et al.} (2018). Finally, work by du Plessis \textit{et al.} (2019), where the RVM (rotating vector model) for pulsars was successfully applied to the AR Sco polarimetric observations, is added evidence for the presence of a rotating magnetic dipole in AR Sco.

 AR Sco was observed in the UV and X-rays with {\it XMM-Newton} by Takata \textit{et al.} (2018), who found orbital modulation, but with no  evidence of eclipses or absorption  features. The  {\it XMM-Newton} data also showed strong X-ray pulsations at the synodic beat frequency $\nu_b$ (=$\nu_s$ - $\nu_o$)  of 8.461100 mHz (P$_b$=118.2s), where $\nu_s$ is the spin frequency at 8.5390 mHz (117.11 s) and $\nu_o$ is the orbital frequency at 0.07792 mHz (3.56 h). These modulations are seen mostly in the soft energy band of 0.15-2.0 keV (Takata \textit{et al.} 2018).  They also reported a very weak signal corresponding to the  side-band frequency of $\nu_s$ + $\nu_o$. The pulse profiles were similar in UV and soft X-rays thus suggestive of similar origin, though the pulse modulation in UV is about 3 times higher than in X-rays.  
 
  The phase averaged X-ray spectrum reported by Takata \textit{et al.} (2018) was best fit by two-temperature (1.1 keV and 8.0 keV) plasma emission models and even better fit by a three-temperature (0.6, 1.7 and 8.0 keV) plasma model with solar abundance, with total flux of 3.2$\times$10$^{-12}$ergs cm$^{-2}$ s$^{-1}$ in the energy band of 0.15 - 12 keV (Takata \textit{et al.} 2018).  The low absorption column density, N$_H$ of 4$\times$10$^{20}$ cm$^{ -2}$ and X-ray luminosity of 4$\times$10$^{30}$ ergs s$^{-1}$  are very different from that seen in intermediate polars.  The X-ray emission likely arises from a non-thermally heated plasma and the system appears to be a WD analogue of a neutron star radio pulsar, although no direct evidence is found for the presence of a non-thermal component in the X-ray spectrum.  Data analyzed in the energy range from 100 MeV to 500 GeV from {\it Fermi} Large Area Telescope (LAT) for the period August 4, 2008 to March 31, 2019, did not lead to any statistically significant detection either (Singh \textit {et al.} 2020).  The exact emission mechanism operating in AR Sco, therefore, continues to be a mystery. 

 Here, we present X-ray observations of AR Sco with the {\it Chandra} X-ray Observatory and the {\it AstroSat} Soft X-ray telescope carried out just a few days apart from each other.
 
\section{Observations}
\subsection{Chandra X-ray Observatory}
 AR Sco was observed with the Advanced CCD Imaging Spectrometer (ACIS-I) onboard the Chandra X-ray Observatory (CXO) from 2017 June 23, 23H:36M:13S to June 24, 07H:25M:59S. The data from this observation (Observation ID: 19711) were downloaded from archives maintained by NASA and analysed with Chandra Interactive Analysis of Observations (CIAO) software version 4.12. Each dataset was reprocessed with the most recently-available calibration (CALDB version 4.9.1 ) applied to it using the $chandra\_repro$ script. The reprocessing created a new level 2 event file for each dataset which was used in all further analyses. The level 2 event files were filtered for Solar Particle flares by analysing their light curves (LCs), and a  useful exposure time of  24790 s was obtained.

\subsection{AstroSat Soft X-ray Telescope}
 {\it AstroSat} (Singh \textit{et al.} 2014) observed AR Sco with the Soft X-ray Telescope (SXT) (Singh \textit{et al.} 2016, 2017) as the prime instrument in the photon counting (PC) mode.  SXT observed the source throughout an orbit of the satellite taking care that the Sun avoidance angle is $\geq$ 45$^\circ$ and RAM angle (the angle between the payload axis to the velocity vector direction of the spacecraft) $>$ 12$^\circ$ to ensure the safety of the mirrors and the detector.  The observation  (Observation ID 9000001350) was started on 2017 June 30 at 06H:36M:51S UT and ended on 2017 Jul 2 at 10H:13M:53S UT.   Level 1 Data from individual orbits were received at the SXT POC (Payload Operation Centre) from the ISSDC (Indian Space Science Data Center). These data  were processed with the sxtpipeline, available in the SXT software (AS1SXTLevel2,  version 1.4b), thus calibrating the source events and extracting Level-2 cleaned event files for the individual orbits. The cleaned event files of all the orbits were then merged into a single cleaned event file using  a $Julia$-based merger tool developed by G. C. Dewangan to avoid the time-overlapping events from the consecutive orbits. The XSELECT (V2.4d) package was used to extract the source spectra and light curves from the processed Level-2 cleaned event files and a useful exposure time of 53610 sec was obtained.  

\section{Analysis and Results}
 An X-ray image of AR Sco obtained with the {\it Chandra} ACIS is shown in Figure 1.  The source lies towards the edge of the field-of-view.  A circular region of 7 arcsec radius was used for extraction of counts from the source.   A box region of size 136"x93" in the vicinity of the source was selected as the background region. 

\begin{figure}[!thbp]
\includegraphics[width=\columnwidth]{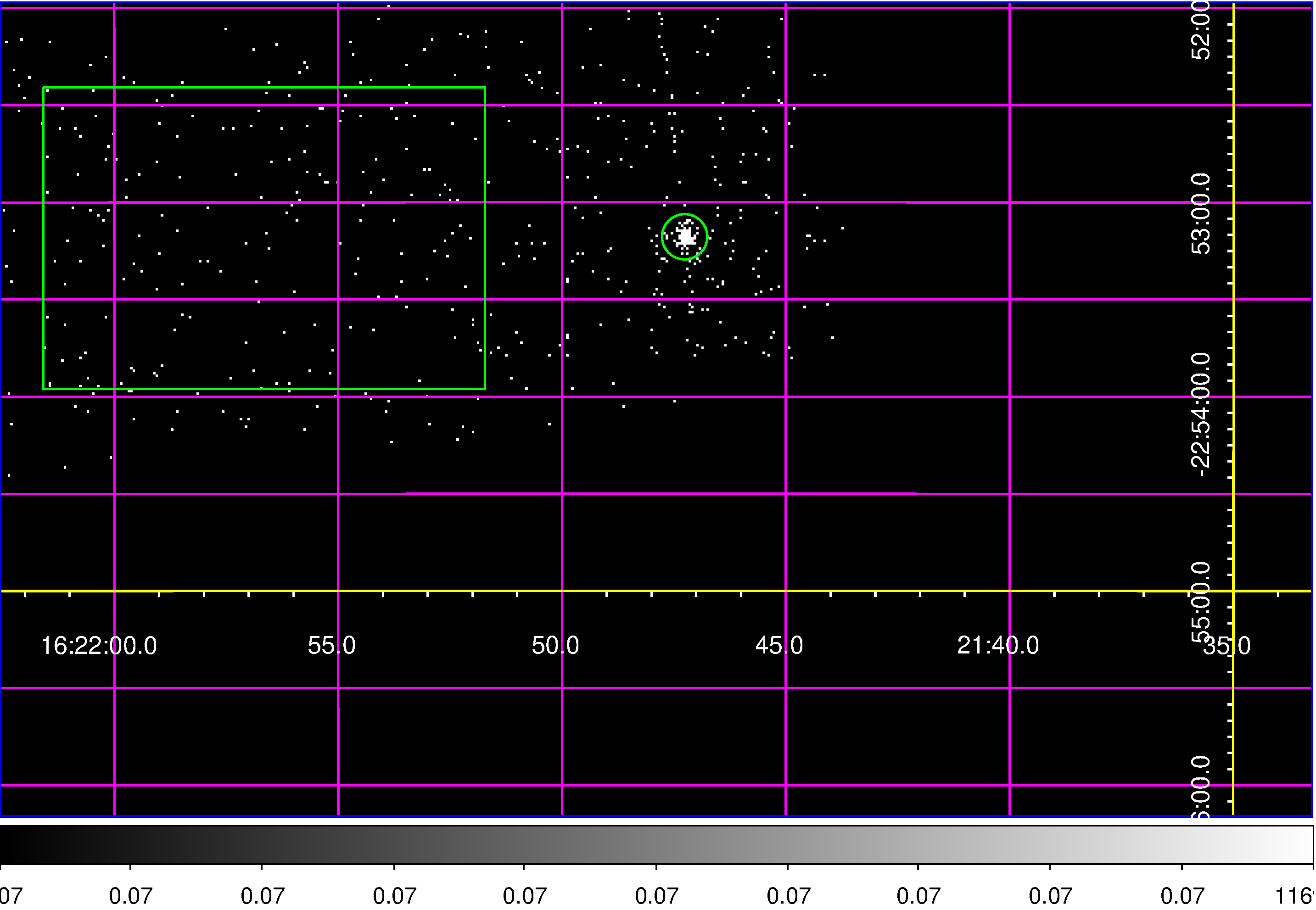}
\caption{{\it Chandra} X-ray image of AR Sco in the energy band of 0.3-7.0 keV showing the source region (circle) and the background region (rectangle) for the extraction of photons used in the analysis.}
\label{figOne}
\end{figure}

An X-ray image of AR Sco taken with the SXT is shown in Figure 2. Source counts were extracted from a circular region of 10 arcmin radius centered on AR Sco.   Both a light curve and a spectrum were extracted from this region for further analysis.

\begin{figure}[!thbp]
\includegraphics[width=\columnwidth]{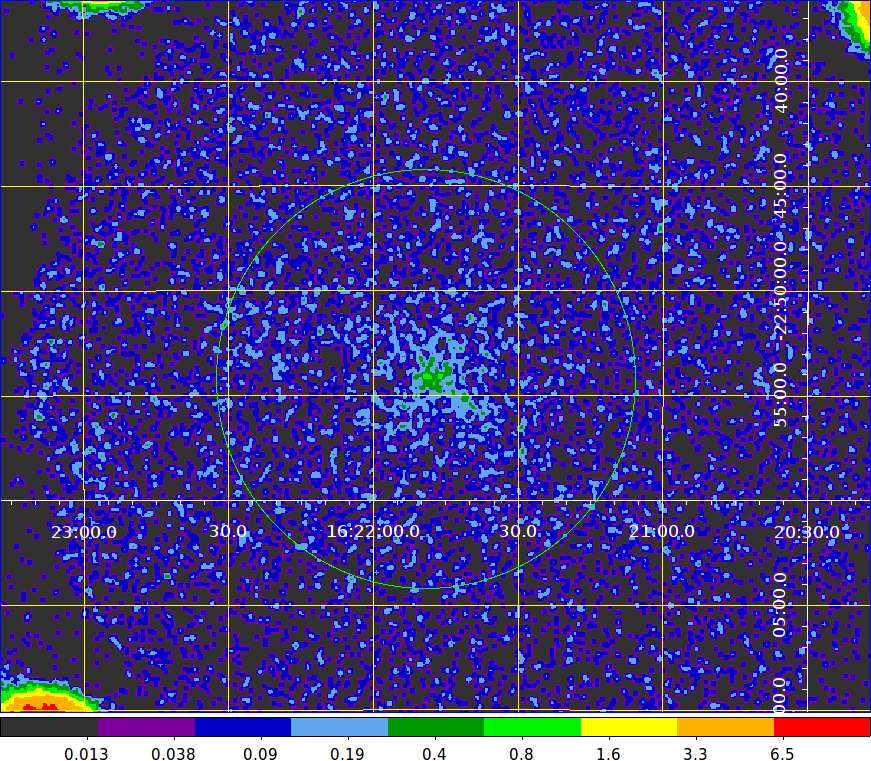}
\caption{{\it AstroSat} SXT image of AR Sco in the energy band of 0.3-7.1 keV showing the circular extraction region of 10 arcmin radius for source counts. Smoothing by a Gaussian with kernel of 3 pixel radius (1 pixel=4 arcsec) has been applied.}
\label{figTwo}
\end{figure}

\subsection{X-ray Light curves}
 The X-ray light curve in the energy range of 0.3 - 7.0 keV obtained from {\it Chandra} observations is shown in Figure 3.  The task 'dmextract' was used to generate the background subtracted source light curve shown.  A binsize of 3.364s was used for extraction. The background region was supplied via the 'dmextract' task parameter 'bkg'.
 
\begin{figure}[!thbp]
\includegraphics[width=0.8\columnwidth, angle=270]{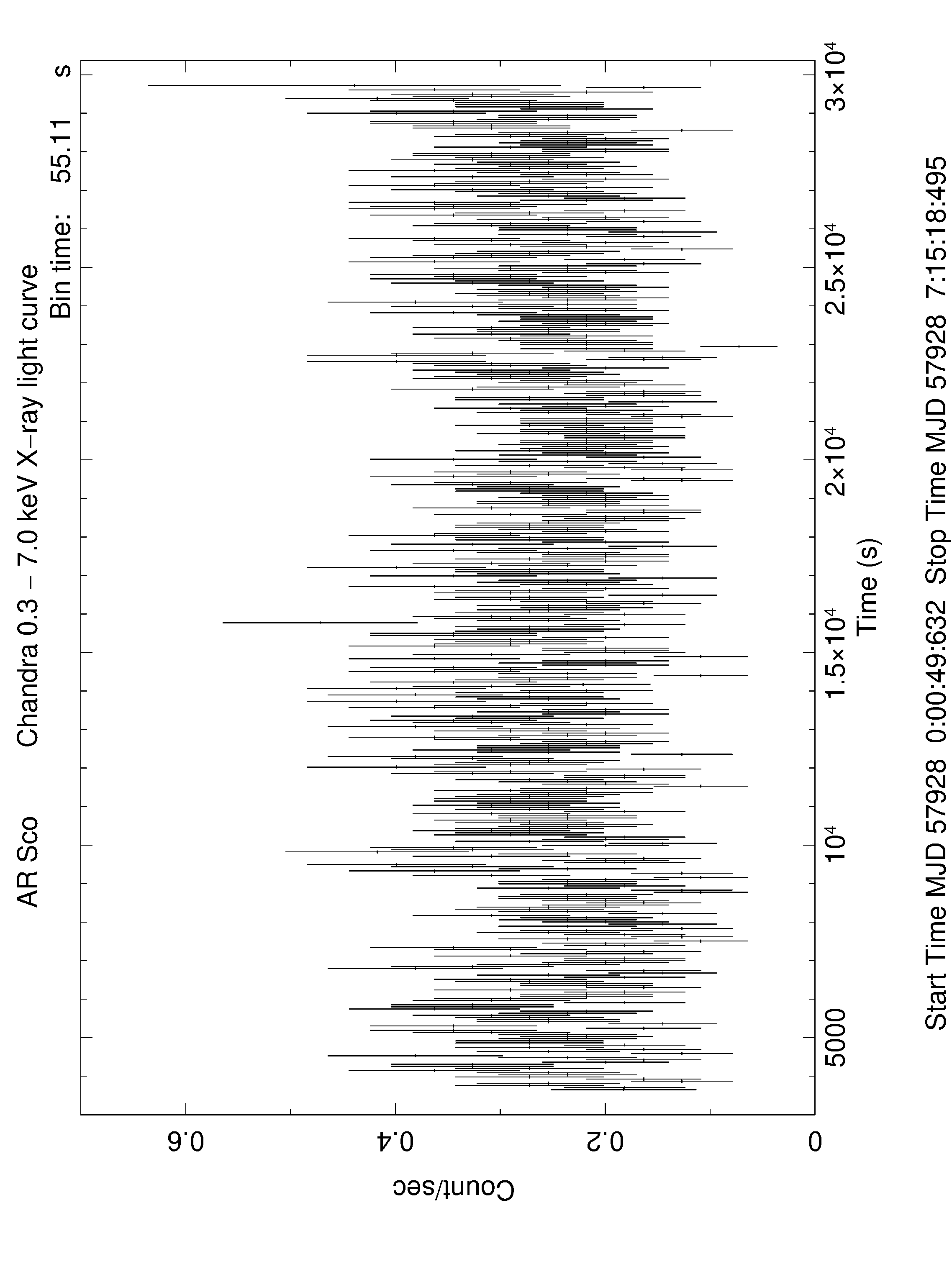}
\caption{X-ray light curve of AR Sco obtained with the {\it Chandra} in the energy range of 0.3-7.0 keV without background subtraction.}
\label{figThree}
\end{figure}

 The SXT light curve of AR Sco was extracted in the useful energy band of 0.3 $-$ 7.1 keV and is shown in  Figure 4. The light curve was extracted with the highest time resolution of 2.3775s.  We also extracted the light curves in the soft energy band of 0.3-2.0 keV from both the $Chandra$ and SXT data, where most of the modulation was seen by Takata \textit{et al.} (2018), and studied their power spectra.

\begin{figure}[!thbp]
\includegraphics[width=0.8\columnwidth, angle=270]{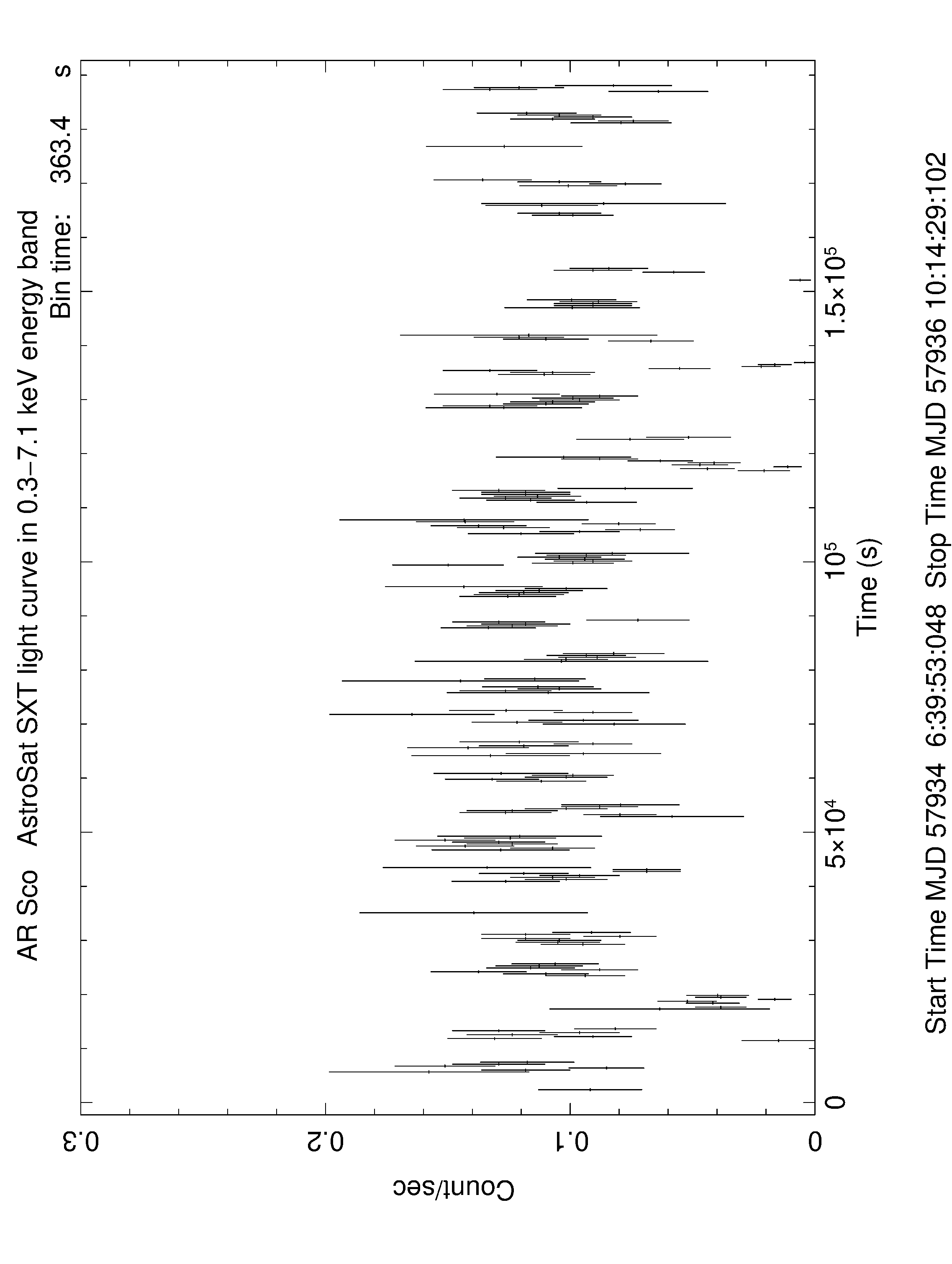}
\caption{X-ray light curve of AR Sco obtained with the {\it AstroSat} SXT in the energy range of 0.3-7.1 keV without any background subtraction.}
\label{figFour}
\end{figure}

\begin{figure}[!thbp]
\includegraphics[width=\columnwidth]{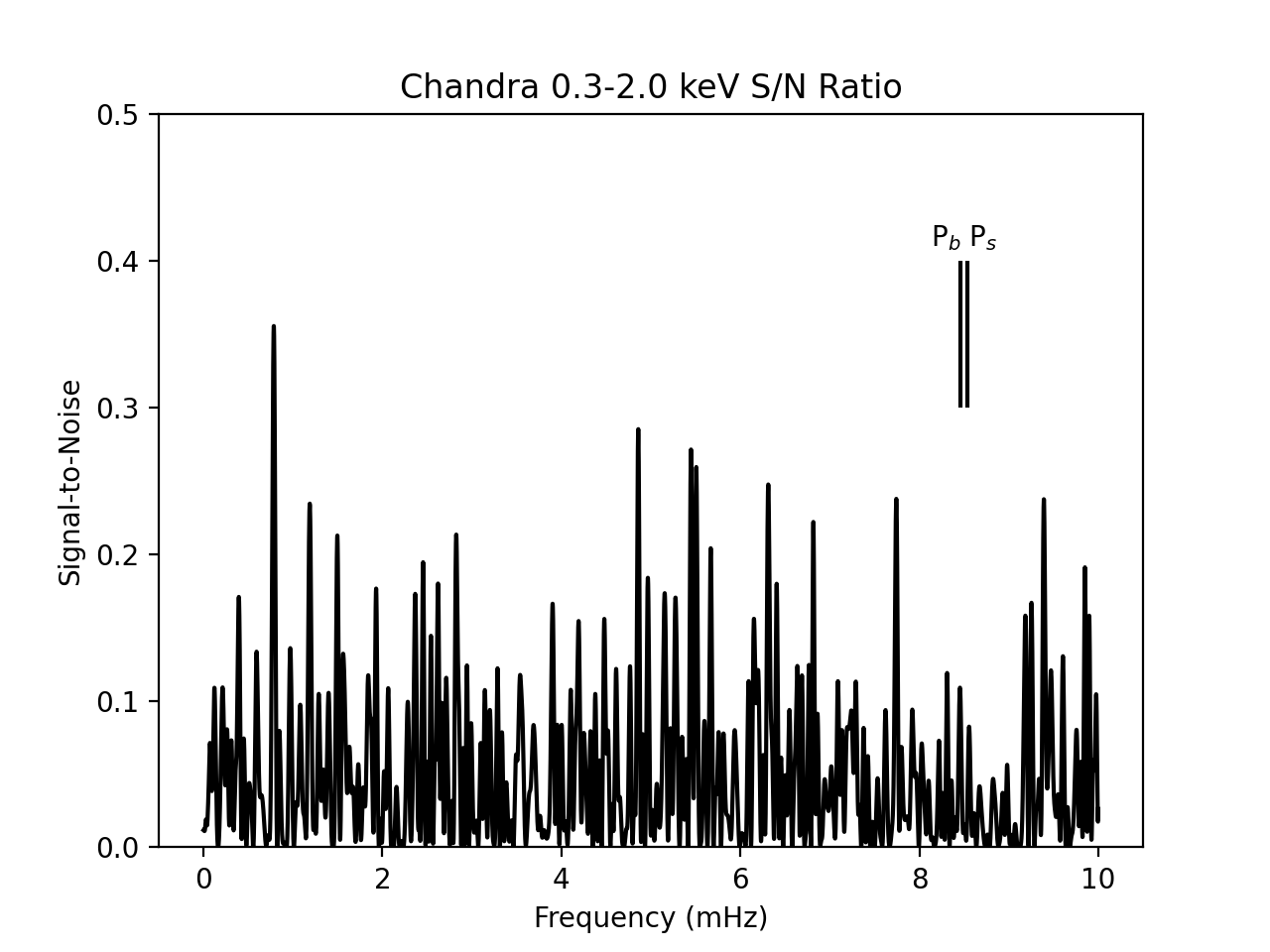}
\caption{Signal-to-Noise ratio at each frequency in mHz for the $Chandra$ 0.3-2.0 keV data, obtained after applying the DFT. The S/N is $<$1 for all frequencies, so the data show no obvious signal at the spin, orbital, or beat periods. The two vertical lines mark the spin and beat frequencies.}
\label{figFive}
\end{figure}

\begin{figure}[!thbp]
\includegraphics[width=\columnwidth]{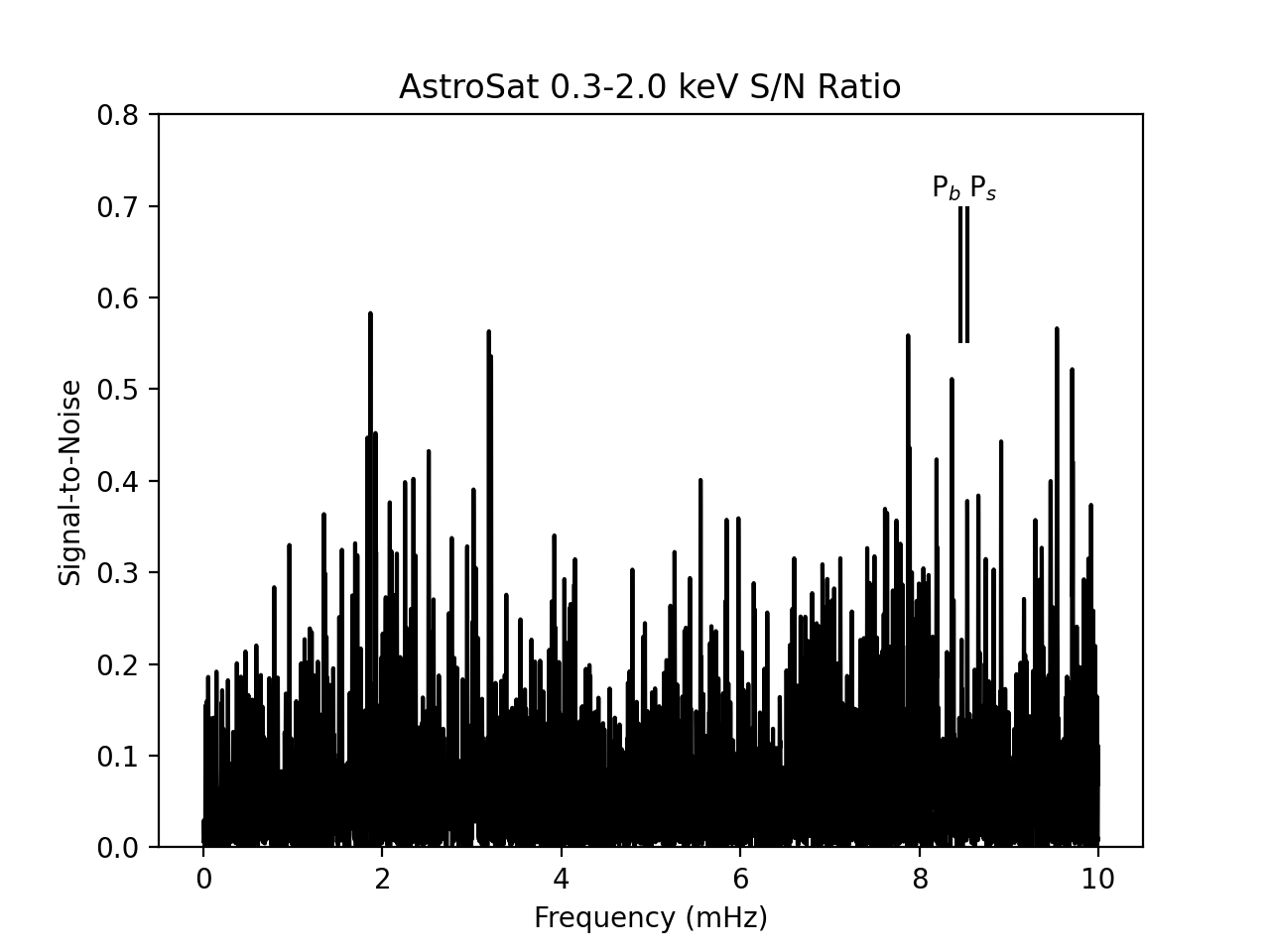}
\caption{Similar to Fig. 5 for SXT 0.3 - 2.0 keV light curve. The S/N is $<$1.5 for all frequencies showing no obvious signal at the spin, orbital, or beat periods. The two vertical lines mark the spin and beat frequencies.}
\label{figSix}
\end{figure}

\begin{figure}[!thbp]
\includegraphics[width=\columnwidth]{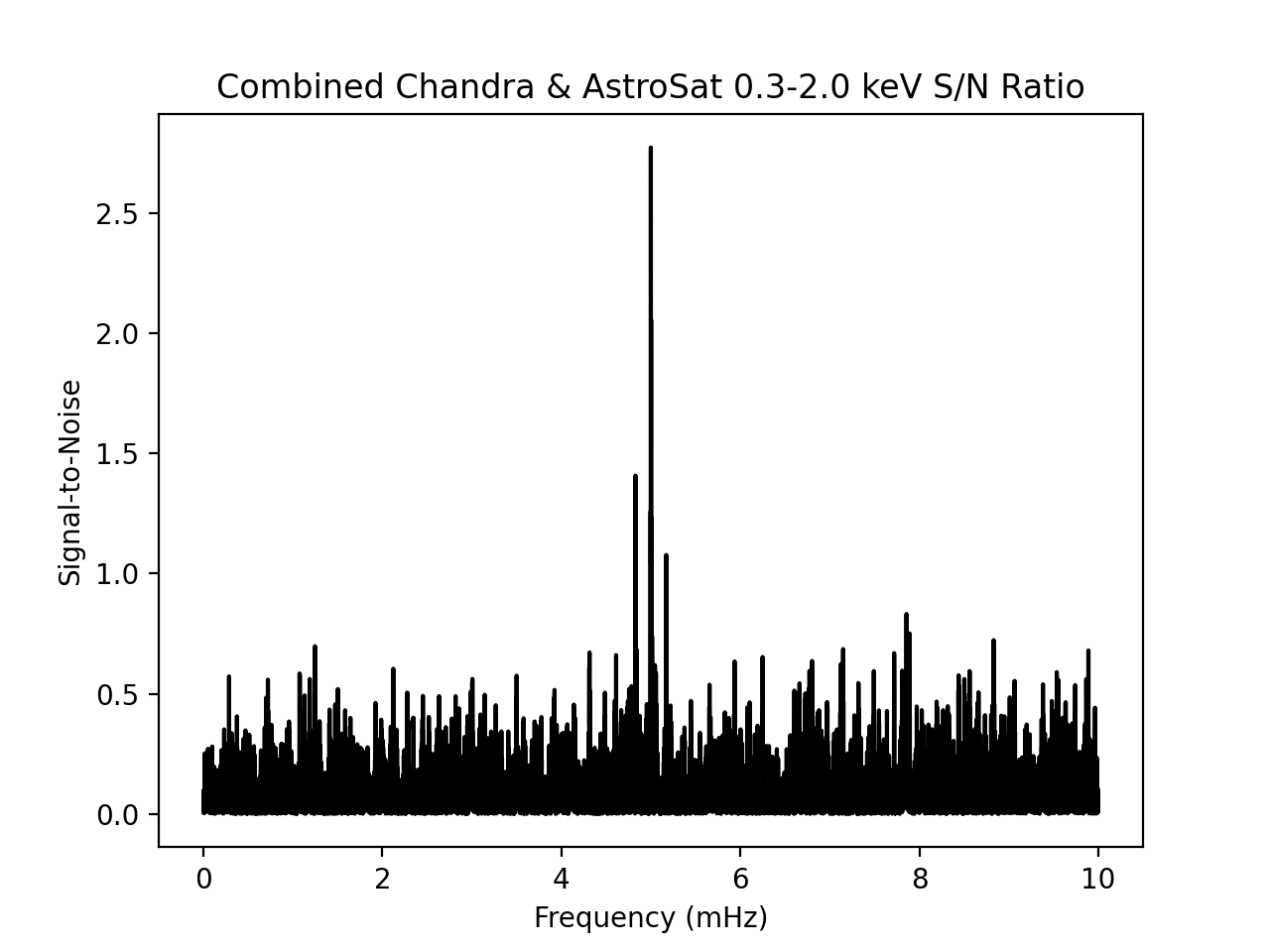}
\caption{Same as above but with the combined $Chandra$ and {\it AstroSat} data injected with a 0.03 counts s$^{-1}$ Poisson sinusoid. This simulation implies that
any modulation of the spin, orbital, and beat periods is less than about 10\% in soft X-rays. }
\label{figSeven}
\end{figure}


 Barycentric corrections are applied to both of the light curves shown in Figs. 3 and 4. The light curves are analyzed using the algorithm of Bretthorst (1988). This algorithm is a formal derivation of the discrete Fourier transform (DFT) using Bayesian statistics and is a generalization of the Lomb-Scargle periodogram \textbf{(see Appendix A)}. The signal-to-noise (S/N) ratio of the resulting power spectra for the 0.3--2.0 keV energy band of {\it Chandra} and {\it AstroSat}  are shown in Figures 5 and 6, respectively. Neither figure shows a signal (or peak) having a S/N ratio of $>5$, or equivalently a significance of $>99$\%, indicating that no signal is detected. Figure 7 shows S/N obtained for the combined {\it Chandra} and {\it AstroSat} light curves that include an injected sinusoidal Poisson signal with an amplitude of  0.03 counts s$^{-1}$ at a frequency of 5 mHz. The S/N ratio of the dominant signal is about 3 or a likelihood of $>$90\%. Note that the S/N ratio decreases quickly from about 3 to 1 between 0.03 to 0.025 counts s$^{-1}$. Based on this simulation, any modulation of the spin, orbital, and beat periods in the combined {\it Chandra} and {\it AstroSat} soft X-ray data are limited to less than about 0.03 counts s$^{-1}$, or a modulation amplitude of less than about 10\%, assuming an average count rate of 0.3 counts s$^{-1}$. Takata \textit{et al.} (2018) report an $\approx$14\% pulse fraction of the beat period in the {\it XMM-Newton} $0.15-0.20$ keV X-ray data and essentially no modulation in the $2-12$ keV data from {\it XMM-Newton}, thus indicating that the pulse fraction decreases rapidly with increasing energy. We conclude that this spectral characteristic combined with the lower energy cutoff of the {\it XMM-Newton} X-ray band compared to that of {\it AstroSat} (0.15 keV vs 0.3 keV), and a small change in the X-ray pulse fraction between the 2016 and 2017 observations, is the reason for the non-detection of the beat period by {\it AstroSat} and {\it Chandra}.

\subsection{X-ray Spectra}
 X-ray spectra were extracted from both the {\it Chandra} and SXT observations. The filtered events file from {\it Chandra} observation was used for extraction of spectra and responses (Ancillary Response Function (ARF)  and spectral  Response Matrix File (RMF)) with the CIAO task 'specextract'. The background region was supplied via the task parameter 'bkgfile'. The parameter 'correctpsf=yes' was used to apply point-source aperture correction to the source ARF file. 

 For the SXT data, a background spectral file ${SkyBkg\_comb\_EL3p5\_Cl\_Rd16p0\_v01.pha}$, derived from a composite of several deep blank sky observations, distributed by the instrument team is used for spectral analysis for all the spectra analysed here. We used  ${sxt\_arf\_excl00\_v04\_20190608.arf}$ as the ARF, and  ${sxt\_pc\_mat\_g0to12.rmf}$ as the RMF for the SXT in this work.  All these files are available at the SXT POC website $https://www.tifr.res.in/~astrosat\_sxt/index.html$.

 The source spectra from both the observations were grouped using the {\it grppha} tool to ensure a minimum of 25 counts per energy bin, prior to further analysis here and below.  The average spectra obtained after background subtraction are shown in Figure 8.  We obtained a net count rate of  0.17 $\pm$ 0.0026 s$^{-1}$ for the {\it Chandra} data and 0.034$\pm$ 0.0012 s$^{-1}$ in the SXT spectrum  in the energy range 0.7$-$7.0 keV, used for a joint spectral analysis. We confined ourselves to the low energy limit of 0.7 keV instead of 0.3 keV as the background subtraction for this very weak source in the SXT indicated almost negligible flux below 0.7 keV while the flux in the {\it Chandra} data at this low energy was slightly higher.  This small discrepancy below 0.7 keV, though not statistically significant,  would have compromised the value of the column density obtained. A line at 6.7 keV is observed clearly (Figure 8) in the X-ray spectrum obtained from the $Chandra$ observations.  The line which is very close to the line expected from ionised Fe XXV indicates thermal emission in the source.

 The two spectra were fitted jointly with a simple plasma emission models {\it apec} using the {\it xspec} program (version 12.9.1; Arnaud (1996) distributed with the heasoft package (version 6.20)). The atomic data base  AtomDB version 3.0.7 ($http://www.atomdb.org$), was used.  We used a common absorber model {\it Tbabs} as a multiplicative model with the model parameter N$_{\rm H}$, i.e., the equivalent Galactic neutral hydrogen column density.  The plasma temperature (kT) and N$_{H}$ parameter were kept free, at least initially.  We used the abundance table  $'aspl'$  given by Asplund \textit{et al.} (2009) for our analysis, and the abundances of all the elements were tied together and could be varied together with respect to the solar values as one parameter.  The normalisation for the plasma component was kept free for the two spectra, and $\chi^2$ minimisation technique was used to find the best fit parameters. The column density, N$_{\rm H}$, was fixed at the low value of 0.1$\times$10$^{21}$ cm$^{-2}$ after it was found to be tending toward zero. This simple model, when fitted to the spectra in the energy range of 0.7 -7.0 keV, gave a good fit with  a plasma temperature $\sim$4 keV and solar elemental abundances.   The best-fit values for this and other models used for fitting are given in Table 1. The normalisation of  the  {\it apec} model for the two sets of data differ only by $\sim$7\%, most probably due to systematic differences in the calibration of the two instruments used.  The total flux obtained for the energy band of 0.5-10 keV is found to be (2.5$\pm$0.8$)\times$10$^{-12}$ ergs cm$^2$ s$^{-1}$.  

 Assuming non-solar abundances for all the elements improved the fit only slightly.  We have explored the possibility of an additional thermal component: a black body or a plasma component at different temperature.  Addition of either of these gave a significant improvement in the $\chi^2$ (and based on the F-statistic values) as can be seen in Table 1,  with the maximum improvement coming from the addition of another thermal component.   The additional blackbody component was best fit for a temperature of 0.13 keV, but it required a significant absorption column density of 1.8-6.7$\times$10$^{21}$cm$^{-2}$ for solar abundances and 2.3-8.2$\times$10$^{21}$cm$^{-2}$ (90\% confidence range) for non-solar abundances.  The best fit to the two-temperature plasma models requires a low temperature component of 0.95$^{+0.20}_{-0.19}$ keV and a high temperature component of 5.0$^{+0.8}_{-0.7}$ keV, elemental abundance of 0.6$\pm$0.3 solar, and a negligible N$_{\rm H}$ in the range of (0.0-2.1)$\times$10$^{21}$cm$^{-2}$.  We also tried two-component $mekal$ plasma models to compare with the results of Takata \textit{et al.} (2018), and found very similar results, except for the low-temperature component which gave the best fit for a temperature of 0.77$^{+0.18}_{-0.16}$, slightly lower but consistent within the errors with that obtained from {\it XMM-Newton} data by Takata \textit{et al.} (2018). 

\vspace{1em}
\begin{table*}[htb]
\tabularfont
\caption{Spectral Parameters obtained from a joint fit to the Chandra and SXT spectra (0.7 - 7.0 keV).}\label{Spectral Paramaters} 
\begin{tabular}{llllllllll}
\topline

Spectral Model&  & & & Parameters & & & & &    \\\midline
 &  N$_H^a$  & kT$_{apec}$ & Z$^b$ & A1$^c$(C,S) & kT$_{bb/apec}$ & A2$^c$(C,S) & $\chi^2_\nu$/dof & Flux$^d_A$ &  Flux$^d_B$\\
&   & keV  &   &     & keV &   & & (C,S) & (C,S) \\\midline
 tbabs*apec  &  0.1 & 4.3$^{+0.4}_{-0.3}$ & 1.0  & 1.55 ,1.66 &  &  & 1.39/274  & 0.97,1.04 & 1.43,1.53 \\
 tbabs*apec  &  0.1 & 4.1$^{+0.4}_{-0.3}$ & 0.6$^{+0.2}_{-0.3}$ & 1.68 ,1.80 &  &   & 1.30/273  & 0.97,1.04  & 1.37,1.47 \\
tbabs*(bbody+apec)  &4.7$^{+2.0}_{-2.9}$ & 4.1$^{+1.1}_{-0.6}$ & 1 & 1.71,1.70 &  0.13$^{+0.03}_{-0.01}$ & 0.05, 0.06 & 1.21/270 & 0.95,1.08 & 1.49,1.46 \\
tbabs*(bbody+apec)  &5.4$^{+2.8}_{-3.1}$ & 4.1$^{+0.8}_{-0.6}$ & 0.7$^{+0.3}_{-0.3}$ & 1.87,1.85 &  0.13$^{+0.02}_{-0.01}$ & 0.07, 0.08 & 1.20/269 & 0.93,1.06 & 1.42,1.39 \\
tbabs*(apec+apec)  &0.2$^{+2.0}_{-0.2}$ & 5.0$^{+0.8}_{-0.7}$ & 1             & 1.5,1.4 & 0.94$^{+0.15}_{-0.18}$ & 0.06, 0.13 & 1.20/270 & 0.97,1.04 & 1.5,1.4 \\
tbabs*(apec+apec)  &0.2$^{+1.9}_{-0.2}$ & 5.0$^{+0.8}_{-0.7}$ & 0.6$^{+0.3}_{-0.3}$ & 1.5,1.4 &  0.95$^{+0.20}_{-0.19}$ & 0.07, 0.15 & 1.20/269 & 0.97,1.04 & 1.5,1.4 \\
tbabs*(mekal+mekal) &  0.3$^{+2.6}_{-0.3}$ & 4.65$^{+0.7}_{-0.7}$ & 0.7$^{+0.2}_{-0.3}$ & 1.44,1.37 &  0.77$^{+0.18}_{-0.16}$ & 0.06, 0.14 & 1.23/269 & 0.99,1.11 & 1.44,1.38\\
\hline
\hline
\end{tabular}
\tablenotes{$^a$ N$_{\rm H}$ is in units of 10$^{21}$ cm$^{-2}$ ; $^b$ Abundance, Z,  is relative to solar values for all the elements; $^c$A1 and A2 are the normalizations for the additive models (as listed) in units of 10$^{-3}$ photons cm$^2$s$^{-1}$; The letters C and S for A1, A2 and for the flux values refer to $Chandra$ and  SXT respectively. $^d$ Fluxes are in units of  10$^{-12}$ ergs cm$^2$ s$^{-1}$ and are quoted for two energy bands: A for 0.5 - 2.0 keV and B for 2.0 - 10 keV; All errors quoted are with 90\% confidence.}
\end{table*}

\begin{figure}[!thbp]
\includegraphics[width=0.74\columnwidth, angle=270]{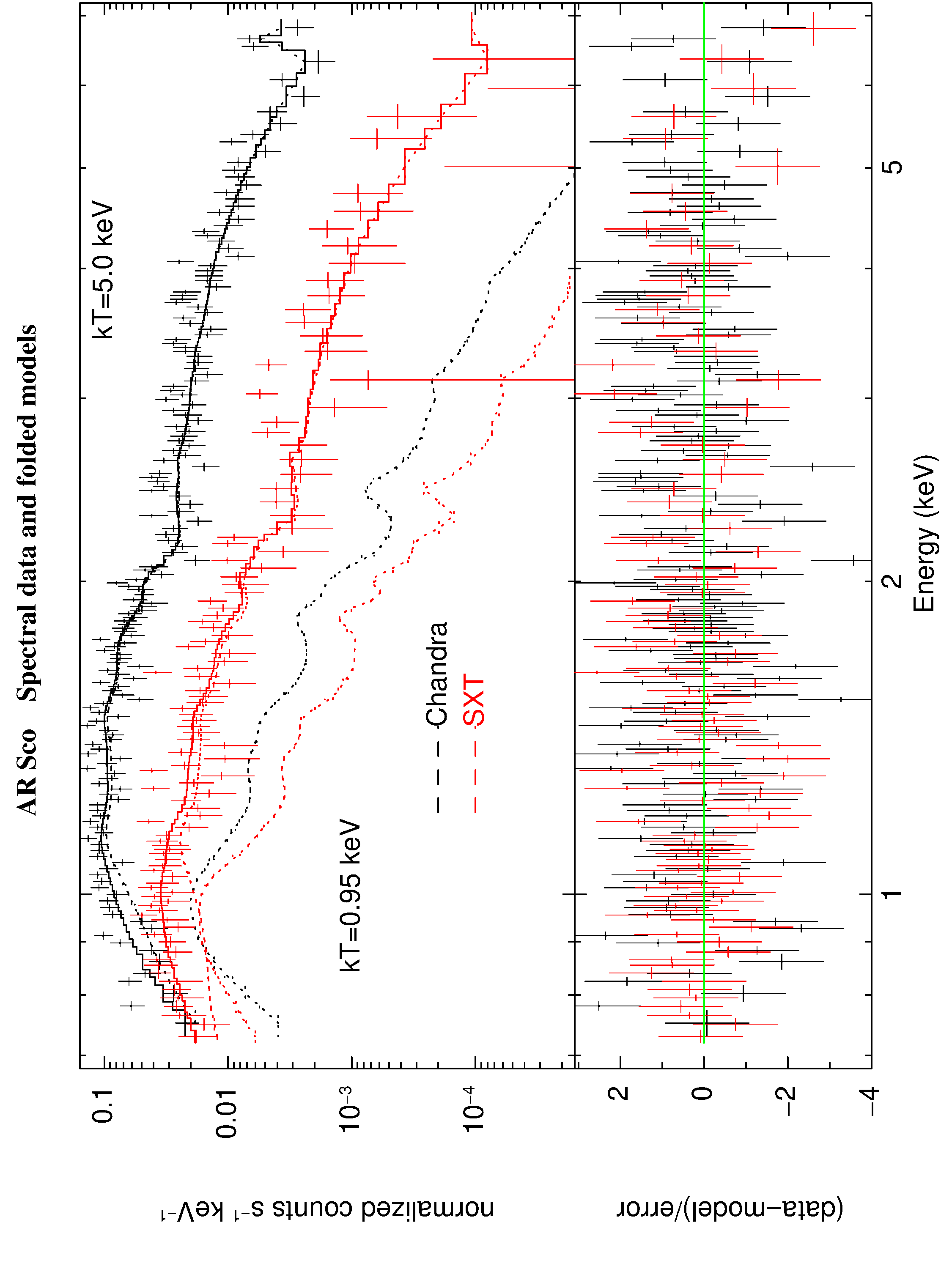}
\caption{X-ray Spectra of AR Sco (top) along with the best fit 2-temperature plasma models ({\it apec}) shown as histograms.  Contribution of each component of the plasma model is shown separately as dotted and dash curves, while the solid line histogram shows the combined contribution from the two components. The rms-normalised residuals per channel are shown in the bottom panel.}
\label{figEight}
\end{figure}

\vspace{-1em}
\section{Discussion and Conclusion}
 Our X-ray observations of AR Sco were carried out in 2017, June-July within a few days of each other with $Chandra$ and {\it AstroSat} SXT, and about nine months after the {\it XMM-Newton} observations (2016 Sept 19).   We find that although on average AR Sco appears about 30\% fainter this is not statistically significant given our errors.   We find that the spectral data from the two instruments can be fit with the same models with remarkable agreement of normalization to within 7\% of each other.   The spectra are clearly thermal as significant emission due to the presence of an ionised iron (Fe XXV) line is seen clearly in the $Chandra$ spectrum.  The Fe line was also reported earlier by Takata \textit{et al.} (2018) in the  {\it XMM-Newton} spectrum.  The best-fit spectral models favour two temperature plasma {\it apec} models with the best fit high  temperature ($\sim$5 keV) component differing from $\sim$8 keV obtained from {\it XMM-Newton} observations.  It should be noted, however, that the SXT has a negligible effective area above the 7.1 keV limit used here, and {\it Chandra} has a very low area above 8.0 keV, and in this particular case the {\it Chandra} data above 7.0 keV were flagged as unusable.  This could be the reason for the difference in the estimated high energy component. The low-temperature component ($\sim$1 keV) and the low column density (negligible) are in agreement, however.

The time series analysis of the {\it Chandra} and {\it AstroSat} soft X-ray data show no periodicities at the WD spin period, the orbital period, or the beat period of the two.  We believe that the lack of such periodicities is due to the lower sensitivity of the {\it Chandra} and {\it AstroSat} X-ray telescopes compared to that of {\it XMM-Newton}.  The XMM light curve contains about 20000 counts, while those of {\it Chandra} and {\it AstroSat} SXT contain about a factor of ten less or about 2000 counts.  Therefore, the relative noise in the {\it Chandra} and {\it AstroSat} periodograms are about a factor of three larger. We also note that the counting statistics are Poissonian because of the low count rates, which may increase the noise in the data caused by quantization error. It is possible that longer exposures by either of these two telescopes may increase the sensitivity enough to detect a modulation of the soft X-ray flux.


\section*{Acknowledgements}
We thank the Indian Space Research Organisation for scheduling the observations and the Indian Space Science Data Centre (ISSDC) for making the data available. This work has been performed utilizing the calibration data-bases and auxillary analysis tools developed, maintained and distributed by AstroSat-SXT team with members from various institutions in India and abroad and the  SXT Payload Operation Center (POC) at the TIFR, Mumbai for the pipeline reduction. The work has also made use of software, and/or web tools obtained from NASA's High Energy Astrophysics Science Archive Research Center (HEASARC), a service of the Goddard Space Flight Center and the Smithsonian Astrophysical Observatory.

\section*{Appendix A}
Arthur Schuster (1905) introduced the periodogram near the beginning of the 20th century and this algorithmic technique has been discussed by many authors since that time; including Lomb (1976) and Scargle (1982) for which the Lomb-Scargle algorithm is known by the astronomical community. The periodogram $C(\omega)$ is defined as the squared magnitude of the discrete Fourier transform of the data $D = {d_1, d_2, ..., d_N}$:
\begin{equation}
    C(\omega) = \frac{1}{N} \Big[R(\omega)^2 + I(\omega)^2\Big] = \frac{1}{N}\Bigg|\sum_{j=1}^N d_j e^{iwt_j}\Bigg|^2 ,
\end{equation}
where
\begin{equation}
    R(\omega) = \sum_{i=1}^N d_i \cos(\omega t_i)
\end{equation}
and
\begin{equation}
    I(\omega) = \sum_{i=1}^N d_i \sin(\omega t_i) .
\end{equation}
A generalization of the periodogram for any time series function was derived using Bayesian principles by Bretthorst in his Ph.D. thesis (1988). The monograph is available for free download at https://bayes.wustl.edu/glb/book.pdf. The following is a brief presentation of Chapters 3 and 4, 'The General Model Equation Plus Noise' In the most general model, the hypothesis H is
\begin{equation}
    H \equiv f(t) = \sum_{j=1}^m B_j G_j(t, \lbrace\omega\rbrace),
\end{equation}
where $f(t)$ is some analytic expression of the time series, $G_j(t, \{\omega\})$ is a particular model function, $B_j$ is the amplitude of the model, and $\{\omega\}$ is a set of parameters (in our case, frequencies). The Bayesian likelihood function for data $D$ is
\begin{equation}
    L(\{B\},\{\omega\},\sigma) \propto \sigma^{-N} \exp \Bigg\lbrace -\frac{NQ}{2\sigma^2} \Bigg\rbrace
\end{equation}
where
\begin{equation}
    Q \equiv \overline{d^2} - \frac{2}{N} \sum_{j=1}^m \sum_{i=1}^N B_j d_i G_j(t_i) + \frac{1}{N} \sum_{j=1}^m \sum_{k=1}^m g_{jk} B_j B_k ,
\end{equation}
\begin{equation}
    g_{jk} = \sum_{i=1}^N G_j(t_i) G_k(t_i) ,
\end{equation}
and $\sigma$ is the standard deviation of the noise.
$Q$ is an $m \times m$ matrix that can be orthogonalized, resulting in a set of eigenvectors of the likelihood equation. Therefore $f(t)$ can be written in terms of a set of orthonormal functions:
\begin{equation}
    f(t) = \sum_{k=1}^m A_k H_k(t) ,
\end{equation}
where
\begin{equation}
    A_k = \sqrt{\lambda_k} \sum_{j=1}^m B_j e_{kj} ,
\end{equation}
\begin{equation}
    B_k = \sum_{j=1}^m \frac{A_j e_{jk}}{\sqrt{\lambda_j}} ,
\end{equation}
and
\begin{equation}
    H_j(t) = \frac{1}{\sqrt{\lambda_j}} \sum_{k=1}^m e_{jk} G_k(t) .
\end{equation}
A change of functions (6) and variables (8) gives the joint likelihood of the new parameters
\small
\begin{equation}
    L(\lbrace A\rbrace,\lbrace\omega\rbrace,\sigma) \propto \sigma^{-N} \exp \Bigg\lbrace -\frac{N}{2\sigma^2} \Bigg[\overline{d^2} - \frac{2}{N} \sum_{j=1}^m A_j h_j + \frac{1}{N} \sum_{j=1}^m A_j^2\Bigg]\Bigg\rbrace ,
\end{equation}
\normalsize
where
\begin{equation}
    h_j \equiv \sum_{i=1}^N d_i H_j(t_i), (1 \leq j \leq m) .
\end{equation}
$h_j$ is the projection of the data onto the orthonormal model function $H_j$. Because we have no knowledge of the amplitudes $A_j$, they need to be marginalized, or integrated out, by performing j integrations, which gives the likelihood
\begin{equation}
    L(\lbrace\omega\rbrace,\sigma) \propto \sigma^{-N+m} \exp\Bigg\lbrace -\frac{N \overline{d^2} - m \overline{h^2}}{2\sigma^2} \Bigg\rbrace ,
\end{equation}
where
\begin{equation}
    \overline{d^2} \equiv \frac{1}{N} \sum_{i=1}^N d_i^2
\end{equation}
and
\begin{equation}
    \overline{h^2} \equiv \frac{1}{m} \sum_{j=1}^m h_j^2 .
\end{equation}
If $\sigma$ is unknown, then it's considered a nuisance parameter and can be eliminated by integrating over all values using the Jeffreys prior $1/\sigma$. This gives an expression for the probability of the form of the "Student t-distribution":
\begin{equation}
    P(\lbrace\omega\rbrace|D,I) \propto \Bigg[1 - \frac{m \overline{h^2}}{N \overline{d^2}}\Bigg]^\frac{m-N}{2} .
\end{equation}

Although $\lbrace A \rbrace$ and $\sigma$ are described as nuisance parameters when deriving the likelihood of the frequency parameters, they are actually of some interest. Bretthorst shows in Chapter 4 that the first posterior moments, i.e., the expected amplitudes, are:
\begin{equation}
    \langle A_j \rangle = h_j ,
\end{equation}
the second posterior moments are:
\begin{equation}
    \langle A_j A_k \rangle = h_j j_k + \sigma^2 \delta_{jk} ,
\end{equation}
the noise variance is:
\begin{equation}
    \langle \sigma^2 \rangle = \frac{1}{N - m -2} \Bigg[ \sum_{i=1}^N d_i^2 - \sum_{j=1}^m h_j^2 \Bigg] ,
\end{equation}
and the signal-to-noise ratio is:
\begin{equation}
    \frac{\rm{Signal}}{\rm{Noise}} = \Bigg\lbrace \frac{m}{N} \Bigg[ 1 + \frac{\overline{h^2}}{\sigma^2} \Bigg] \Bigg\rbrace^{\frac{1}{2}}
\end{equation}

\end{document}